# Deep-sea deployment of the KM3NeT neutrino telescope detection units by self-unrolling

S. Aiello[a], A. Albert[bd,b], S. Alves Garre[c], Z. Aly[d], F. Ameli[e], E.G. Anassontzis[f], M. Andre[g], G. Androulakis[h], M. Anghinolfi[i], M. Anguita[j], G. Anton[k], M. Ardid[l], J. Aublin[m], C. Bagatelas[h], R. Bakker[ar], G. Barbarino[o,p], B. Baret[m], S. Basegmez du Pree[n], M. Bendahman[q], E. Berbee[n,*], A.M. van den Berg[r], V. Bertin[d], S. Biagi[s], M. Billault[d], M. Bissinger[k], M. Boettcher[t], J. Boumaaza[q], M. Bouta[u], M. Bouwhuis[n], C. Bozza[v], H.Brânzaş[w], R. Bruijn[n,x], J. Brunner[d], E. Buis[y], R. Buompane[o,z], J. Busto[d], G. Cacopardo[s], B. Caiffi[i], L. Caillat[d], D. Calvo[c], A. Capone[aa,e], V. Carretero[c], P. Castaldi[ab,ac], S. Celli[aa,e,be], M. Chabab[ad], N. Chau[m], A. Chen[ae], S. Cherubini[s,af], V. Chiarella[ag], T. Chiarusi[ab], M. Circella[ah], R. Cocimano[s], J.A.B. Coelho[m], A. Coleiro[m], M. Colomer Molla[m,c], S. Colonges[m], R. Coniglione[s], I. Corredoira[c], A. Cosquer[d], P. Coyle[d], A. Creusot[m], G. Cuttone[s], C. D'Amato[s], A. D'Onofrio[o,z], R. Dallier[ai], M. De Palma[ah,aj], I. Di Palma[aa,e], A.F. Díaz[j], D. Diego-Tortosa[l], C. Distefano[s], A. Domi[i,d,ak], R. Donà[ab,al], C. Donzaud[m], D. Dornic[d], M. Dörr[am], D. Drouhin[bd,b], T. Eberl[k], A. Eddyamoui[q], T. van Eeden[n], D. van Eijk[n], I. El Bojaddaini[u], D. Elsaesser[am], A. Enzenhöfer[d], V. Espinosa[l], P. Fermani[aa,e], G. Ferrara[s,af], M.D. Filipović[an], F. Filippini[ab,al], L.A. Fusco[d], O. Gabella[ao], T. Gal[k], A. Garcia Soto[n], F. Garufi[o,p], Y. Gatelet[m], N. Geißelbrecht[k], L. Gialanella[o,z], E. Giorgio[s], L. Gostiaux[ar,zz], S.R. Gozzini[k], R. Gracia[n], K. Graf[k], D. Grasso[ap], G. Grella[aq], A. Grmek[s], D. Guderian[bf], C. Guidi[i,ak], S. Hallmann[k], H. Hamdaoui[q], H. van Haren[ar,*], J. van Heerwaarden[ar], A. Heijboer[n], A. Hekalo[am], S. Henry[d], J.J. Hernández-Rey[c], T. Hillebrand[ar], J. Hofestädt[k], F. Huang[as], W. Idrissi Ibnsalih[o,z], A. Ilioni[m], G. Illuminati[m,c], C.W. James[at], M. de Jong[n], P. de Jong[n,x], B.J. Jung[n], M. Kadler[am], P. Kalaczyński[au], O. Kalekin[k], U.F. Katz[k], N.R. Khan Chowdhury[c], G. Kistauri[av], F. van der Knaap[y], E.N. Koffeman[n,x], P. Kooijman[x,bg], A. Kouchner[m,aw], M. Kreter[t], V. Kulikovskiy[i], M. Laan[ar], R. Lahmann[k], P. Lamare[d], G. Larosa[s], J. Laurence[d], R. Le Breton[m], O. Leonardi[s], F. Leone[s,af], E. Leonora[a], G. Levi[ab,al], M. Lincetto[d], M. Lindsey Clark[m], T. Lipreau[ai], F. Longhitano[a], D. Lopez-Coto[ax], L. Maderer[m], J. Mańczak[c], K. Mannheim[am], A. Margiotta[ab,al], A. Marinelli[o], C. Markou[h], L. Martin[ai], J.A. Martínez-Mora[l], A. Martini[ag], F. Marzaioli[o,z], S. Mastroianni[o], S. Mazzou[ad], K.W. Melis[n], G. Miele[o,p], P. Migliozzi[o], E. Migneco[s], P. Mijakowski[au], L.S. Miranda[ay], C.M. Mollo[o], M. Mongelli[ah], M. Morganti[ap,bh], M. Moser[k], A. Moussa[u], R. Muller[n], D. Muñoz Pérez[c], M. Musumeci[s], L. Nauta[n], S. Navas[ax], C.A. Nicolau[e], B. Ó Fearraigh[n,x], M. O'Sullivan[at], M. Organokov[as], A. Orlando[s], J. Palacios González[c], G. Papalashvili[av], R. Papaleo[s], C. Pastore[ah], A.M. Păun[w], G.E. Păvălaş[w], C. Pellegrino[al,bi], M. Perrin-Terrin[d], P. Piattelli[s], C. Pieterse[c], K. Pikounis[h], O. Pisanti[o,p], C. Poirè[l], V. Popa[w], T. Pradier[as], G. Pühlhofer[az], S. Pulvirenti[s], O. Rabyang[t], F. Raffaelli[ap], N. Randazzo[a], S. Razzaque[ay], D. Real[c], S. Reck[k], G. Riccobene[s], M. Richer[as], S. Rivoire[ao], A. Rovelli[s], F. Salesa Greus[c], D.F.E. Samtleben[n,ba], A. Sánchez Losa[ah], M. Sanguineti[i,ak], A. Santangelo[az], D. Santonocito[s], P. Sapienza[s], J. Schnabel[k], J. Schumann[k], J. Seneca[n], I. Sgura[ah], R. Shanidze[av], A. Sharma[bb], F. Simeone[e], A. Sinopoulou[h], B. Spisso[aq,o], M. Spurio[ab,al], D. Stavropoulos[h], J. Steijger[n], S.M. Stellacci[aq,o], M. Taiuti[i,ak], Y. Tayalati[q], E. Tenllado[ax], D. Tézier[d], T. Thakore[c], S. Tingay[at], E. Tzamariudaki[h], D. Tzanetatos[h], V. Van Elewyck[m,aw], G. Vasileiadis[ao], F. Versari[ab,al], S. Viola[s], D. Vivolo[o,p], G. de Wasseige[m], J. Wilms[bc], R. Wojaczyński[au], E. de Wolf[n,x,*], S. Zavatarelli[i], A. Zegarelli[aa,e], D. Zito[s], J.D. Zornoza[c], J. Zúñiga[c], N. Zywucka[t], (The KM3NeT Collaboration)

*corresponding author
Email addresses: eberbee@km3net.de (E. Berbee), hans.van.haren@nioz.nl (H. van Haren), e.dewolf@nikhef.nl (E. de Wolf)

[a] INFN, Sezione di Catania, Via Santa Sofia 64, Catania, 95123 Italy
[b] IN2P3, IPHC, 23 rue du Loess, Strasbourg, 67037 France
[c] IFIC - Instituto de Física Corpuscular (CSIC - Universitat de València), c/Catedrático José Beltrán, 2, 46980 Paterna, Valencia, Spain
[d] Aix Marseille Univ, CNRS/IN2P3, CPPM, Marseille, France
[e] INFN, Sezione di Roma, Piazzale Aldo Moro 2, Roma, 00185 Italy
[f] Physics Department, N. and K. University of Athens, Athens, Greece
[g] Universitat Politècnica de Catalunya, Laboratori d'Aplicacions Bioacústiques, Centre Tecnològic de Vilanova i la Geltrú, Avda. Rambla Exposició, s/n, Vilanova i la Geltrú, 08800 Spain
[h] NCSR Demokritos, Institute of Nuclear and Particle Physics, Ag. Paraskevi Attikis, Athens, 15310 Greece
[i] INFN, Sezione di Genova, Via Dodecaneso 33, Genova, 16146 Italy
[j] University of Granada, Dept. of Computer Architecture and Technology/CITIC, 18071 Granada, Spain
[k] Friedrich-Alexander-Universität Erlangen-Nürnberg, Erlangen Centre for Astroparticle Physics, Erwin-Rommel-Straße 1, 91058 Erlangen, Germany
[l] Universitat Politècnica de València, Instituto de Investigación para la Gestión Integrada de las Zonas Costeras, C/ Paranimf, 1, Gandia, 46730 Spain
[m] Université de Paris, CNRS, Astroparticule et Cosmologie, F-75013 Paris, France
[n] INFN, Sezione di Napoli, Complesso Universitario di Monte S. Angelo, Via Cintia ed. G, Napoli, 80126 Italy
[o] Università di Napoli “Federico II”, Dip. Scienze Fisiche “E. Pancini”, Complesso Universitario di Monte S. Angelo, Via Cintia ed. G, Napoli, 80126 Italy
[p] Nikhef, National Institute for Subatomic Physics, PO Box 41882, Amsterdam, 1009 DB Netherlands
[q] University Mohammed V in Rabat, Faculty of Sciences, 4 av. Ibn Battouta, B.P. 1014, R.P. 10000 Rabat, Morocco
[r] KVI-CART University of Groningen, Groningen, the Netherlands
[s] INFN, Laboratori Nazionali del Sud, Via S. Sofia 62, Catania, 95123 Italy
[t] North-West University, Centre for Space Research, Private Bag X6001, Potchefstroom, 2520 South Africa
[u] University Mohammed I, Faculty of Sciences, BV Mohammed VI, B.P. 717, R.P. 60000 Oujda, Morocco
[v] Università di Salerno e INFN Gruppo Collegato di Salerno, Dipartimento di Matematica, Via Giovanni Paolo II 132, Fisciano, 84084 Italy
[w] ISS, Atomistilor 409, Măgurele, RO-077125 Romania
[x] University of Amsterdam, Institute of Physics/IHEF, PO Box 94216, Amsterdam, 1090 GE Netherlands
[y] TNO, Technical Sciences, PO Box 155, Delft, 2600 AD Netherlands
[z] Università degli Studi della Campania “Luigi Vanvitelli”, Dipartimento di Matematica e Fisica, viale Lincoln 5, Caserta, 81100 Italy
[aa] Università La Sapienza, Dipartimento di Fisica, Piazzale Aldo Moro 2, Roma, 00185 Italy
[ab] INFN, Sezione di Bologna, v.le C. Berti-Pichat, 6/2, Bologna, 40127 Italy
[ac] Università di Bologna, Dipartimento di Ingegneria dell'Energia Elettrica e dell'Informazione “Guglielmo Marconi”, v.le C. Berti-Pichat, 6/2, Bologna, 40127 Italy
[ad] Cadi Ayyad University, Physics Department, Faculty of Science Semlalia, Av. My Abdellah, P.O.B. 2390, Marrakech, 40000 Morocco

[ae] University of the Witwatersrand, School of Physics, Private Bag 3, Johannesburg, Wits 2050 South Africa
[af] Università di Catania, Dipartimento di Fisica e Astronomia "Ettore Majorana", Via Santa Sofia 64, Catania, 95123 Italy
[ag] INFN, LNF, Via Enrico Fermi, 40, Frascati, 00044 Italy
[ah] INFN, Sezione di Bari, Via Amendola 173, Bari, 70126 Italy
[ai] Subatech, IMT Atlantique, IN2P3-CNRS, Université de Nantes, 4 rue Alfred Kastler - La Chantrerie, Nantes, BP 20722 44307 France
[aj] University of Bari, Via Amendola 173, Bari, 70126 Italy
[ak] Università di Genova, Via Dodecaneso 33, Genova, 16146 Italy
[al] Università di Bologna, Dipartimento di Fisica e Astronomia, v.le C. Berti-Pichat, 6/2, Bologna, 40127 Italy
[am] University Würzburg, Emil-Fischer-Straße 31, Würzburg, 97074 Germany
[an] Western Sydney University, School of Computing, Engineering and Mathematics, Locked Bag 1797, Penrith, NSW 2751 Australia
[ao] Laboratoire Univers et Particules de Montpellier, Place Eugène Bataillon - CC 72, Montpellier Cédex 05, 34095 France
[ap] INFN, Sezione di Pisa, Largo Bruno Pontecorvo 3, Pisa, 56127 Italy
[aq] Università di Salerno e INFN Gruppo Collegato di Salerno, Dipartimento di Fisica, Via Giovanni Paolo II 132, Fisciano, 84084 Italy
[ar] NIOZ (Royal Netherlands Institute for Sea Research) and Utrecht University, PO Box 59, Den Burg, Texel, 1790 AB, the Netherlands
[as] Université de Strasbourg, CNRS IPHC UMR 7178, 23 rue du Loess, Strasbourg, 67037 France
[at] International Centre for Radio Astronomy Research, Curtin University, Bentley, WA 6102, Australia
[au] National Centre for Nuclear Research, 02-093 Warsaw, Poland
[av] Tbilisi State University, Department of Physics, 3, Chavchavadze Ave., Tbilisi, 0179 Georgia
[aw] Institut Universitaire de France, 1 rue Descartes, Paris, 75005 France
[ax] University of Granada, Dpto. de Física Teórica y del Cosmos & C.A.F.P.E., 18071 Granada, Spain
[ay] University of Johannesburg, Department Physics, PO Box 524, Auckland Park, 2006 South Africa
[az] Eberhard Karls Universität Tübingen, Institut für Astronomie und Astrophysik, Sand 1, Tübingen, 72076 Germany
[ba] Leiden University, Leiden Institute of Physics, PO Box 9504, Leiden, 2300 RA Netherlands
[bd] Università di Pisa, Dipartimento di Fisica, Largo Bruno Pontecorvo 3, Pisa, 56127 Italy
[bc] Friedrich-Alexander-Universität Erlangen-Nürnberg, Remeis Sternwarte, Sternwartstraße 7, 96049 Bamberg, Germany
[bd] Université de Strasbourg, Université de Haute Alsace, GRPHE, 34, Rue du Grillenbreit, Colmar, 68008 France
[be] Gran Sasso Science Institute, GSSI, Viale Francesco Crispi 7, L'Aquila, 67100 Italy
[bf] University of Münster, Institut für Kernphysik, Wilhelm-Klemm-Str. 9, Münster, 48149 Germany
[bg] Utrecht University, Department of Physics and Astronomy, PO Box 80000, Utrecht, 3508 TA Netherlands
[bh] Accademia Navale di Livorno, Viale Italia 72, Livorno, 57100 Italy
[bi] INFN, CNAF, v.le C. Berti-Pichat, 6/2, Bologna, 40127 Italy
[bj] NRC "Kurchatov Institute", A.I. Alikhanov Institute for Theoretical and Experimental Physics, Bolshaya Cheremushkinskaya ulitsa 25, Moscow, 117218 Russia

[zz] Presently at: Laboratoire de Mécanique des Fluides et d'Acoustique,UMRCNRS5509, École Centrale de Lyon, Université de Lyon, 36 avenue Guy de Collongue, 69134 Écully cedex, France

ABSTRACT

**KM3NeT is a research infrastructure being installed in the deep Mediterranean Sea. It will house a neutrino telescope comprising hundreds of networked moorings – detection units or strings – equipped with optical instrumentation to detect the Cherenkov radiation generated by charged particles from neutrino-induced collisions in its vicinity. In comparison to moorings typically used for oceanography, several key features of the KM3NeT string are different: the instrumentation is contained in transparent and thus unprotected glass spheres; two thin Dyneema® ropes are used as strength members; and a thin delicate backbone tube with fibre-optics and copper wires for data and power transmission, respectively, runs along the full length of the mooring. Also, compared to other neutrino telescopes such as ANTARES in the Mediterranean Sea and GVD in Lake Baikal, the KM3NeT strings are more slender to minimise the amount of material used for support of the optical sensors. Moreover, the rate of deploying a large number of strings in a period of a few years is unprecedented. For all these reasons, for the installation of the KM3NeT strings, a custom-made, fast deployment method was designed. Despite the length of several hundreds of metres, the slim design of the string allows it to be compacted into a small, re-usable spherical launching vehicle instead of deploying the mooring weight down from a surface vessel. After being lowered to the seafloor, the string unfurls to its full length with the buoyant launching vehicle rolling along the two ropes. The design of the vehicle, the loading with a string, and its underwater self-unrolling are detailed in this paper.**

# 1 Introduction

The routine method for oceanographers to make observations at a fixed position in the sea as a function of time is by mooring a string of instruments, see for example [1, 2]. A sufficiently weighted anchor holds position under current and/or wave drag, while the string is held upright by buoyancy elements. The top buoyancy may be at the sea surface, which requires a somewhat loosely tethered string because of the necessary slack to allow for wave action. For precise measurements of flow past a fixed point in space and the study of its dynamical processes, one requires minimal instrumentation movement, as well as a top buoy below the surface [1]. Underwater top buoys need to be sufficiently deep, well below the surf zone and preferably in low current flow-speed areas, whilst the design of the string needs to be optimised for minimal drag and sufficient tension in the cable. Mooring cables of various make are used, such as coated/anodized steel, Kevlar or nylon. An entirely submersed mooring without surface markers requires a release mechanism for recovery of the string to the surface leaving the anchor weight at the seafloor.

In oceanography, most often stand-alone 'self-contained' instrumentation is used with electronics including power-supply (batteries) and data storage held in a corrosion-proof and pressure-resistant container. Every 10 m depth increase adds approximately 1 bar ($10^5$ N $m^{-2}$ ~ 1 atm) of static environmental pressure, and salt water is highly corrosive for modern electronics. Typical containers are made of seawater resistant steel, titanium, or glass that is protected by a sturdy plastic housing for overboard and handling operations. The self-contained set-up limits the amount of data sampled, depending on the duration of the mooring period.

The standard deployment procedure is to put the top buoy at the sea surface, unroll the cable from a winch, add instruments and finally drop the anchor-frame in 'free fall', i.e. without further control from the deployment ship. Depending on the sea-current conditions and on the length of the mooring, the precision of horizontal positioning at the seafloor is about 50 m per 1000 m water depth increment. Precise location of the anchor to within ±10 m is determined afterwards from triangulation measurements when an acoustic (release) beacon is included at the bottom of the string.

More controlled deployment with weight being lowered down first is seldom done, also because the strain on the mooring cable considerably increases due to ship motion upon surface wave action. The imprecise 'free fall' method described above can be performed under much worse sea state conditions and with thinner mooring cables as strength members. For oceanographic purposes, multiple moorings located to within 100 m from each other in water depths exceeding 1000 m are extremely rare in practice. The practice limits the observational view as spatially one-dimensional (1D) in the vertical, and in time. The reason is usually a matter of costs and the wish to cover as much as possible of the vast ocean, rather than detailing the small-scale 3D ocean dynamics.

For a deep-sea observatory connected to a shore station for continuous data taking during a considerable number of years, the approach has to be different. During the design phase of the KM3NeT cubic kilometre sized underwater neutrino telescope detector (Fig. 1) [3], technical and budgetary requirements and constraints were formulated for the deployment method of the instrumentation.

Several requirements are different from those of standard oceanographic moorings [4], and include: (i) a 3D viewpoint in contrast to the routine oceanographic practice of the deployment of a single 1D string; (ii) minimal protection of the glass spheres containing photomultiplier tubes (PMT)s, which should not be covered by anything – in particular not by full protective plastic caps typically used by oceanographers; (iii) the delicacy of the electro-optical backbone cable running along the full length of the string; (iv) a mooring design minimising current-flow drag with thin strength-members as well as data and power transport cables; (v) the expected lifetime of the mooring of nominally 10, preferably 15 years under water; and (vi) the demand of deploying multiple moorings in a relatively short time span to avoid prohibitively expensive sea operations both in time and money.

KM3NeT's predecessor ANTARES consists of 12 lines 60 m apart horizontally holding a total of 885 optical sensors [5] so that the detector comprised a volume of about 0.01 $km^3$.

Deployment of the lines was weight down, because of the demanded precision for line-positioning smaller than a few metres. As a result, sea operations could only be performed under less than 5 m $s^{-1}$ wind speeds, and rather bulky strength-members were used that could safely hold 70 kN dynamical tension. Each optical sensor was held in a 0.44 m diameter glass sphere, with three glass spheres being mounted in a titanium frame including a separate titanium housing containing its main electronics. For more than ten years, the ANTARES telescope has demonstrated successfully the feasibility of building and operating a deep-sea neutrino telescope. However, scaling up its technology to the size of KM3NeT appeared to be costly. Moreover, a modified design of the optical module improves the sensitivity of the telescope significantly.

For KM3NeT, the optical module design was modified, from a glass sphere comprising a single large PMT without main electronics to one with the same diameter but housing 31 small PMTs, calibration devices and the full front-end and readout electronics [3]. This new optical module design, together with the requirement of cost reduction forced a redesign of the entire deployment procedure, starting with a slender frame to hold an optical module in an unprotected glass sphere so as to obscure less than 5% of the PMTs field of view [6]. The new design also demanded the development of a new deployment method, of which the test version has been presented in Ref. [7].

In this paper, we report on the finalised specific deployment design using a launching vehicle, dubbed LOM for Launcher of Optical Modules, which is custom-made for the new detection unit 'DU-string' [6] of the underwater neutrino telescope of which the first DU-strings are now successfully operational [8]. We also discuss particular tools that were additionally custom-made to use the deployment design.

## 2 The KM3NeT detection unit and optical module

The KM3NeT design of an optical module is a 0.44 m diameter $6\times10^{7}$ N $m^{-2}$ pressure-resistant Vitrovex glass sphere providing +0.25 kN buoyancy. The glass sphere is equipped with 31 3-inch PMTs and the associated electronics, sensors and devices for data acquisition, monitoring, calibration and long-range communication with shore. The optical modules are enclosed in slender titanium collars which are attached with two polyethylene PE bollards in predefined positions to two 0.004 m diameter, 12 kN strength approximately neutrally buoyant synthetic Dyneema® ropes [6, 7]. The ropes link the anchor-frame, resting on the seafloor, in a string to 18 optical modules and a 1.35 kN syntactic foam top buoy, and provide support to the backbone of a 0.007 m diameter oil-filled PE tube – dubbed VEOC (Vertical Electro Optical Cable) - guiding optical fibres and copper wires for data and power transport (Fig. 2). The fibres and wires in the VEOC connect to the optical modules via an oil-filled break-out box.

The above described ensemble is dubbed DU-string. The total net buoyancy is about 3.5 kN and the maximum horizontal deflection of the top buoy of a 700 m long DU-string under 0.15 m $s^{-1}$ peak current-flow speeds is about 100 m. The typical flow speed at the KM3NeT sites is between 0.05 and 0.1 m $s^{-1}$. The pair of Dyneema ropes are pre-stretched (2.5 kN) and marked under tension (2 kN) at approximately 2 m intervals to within centimeters precision for the mounting of the optical modules. In case of a height difference in mounting of the two sides of an optical module in the DU-string, the self-correcting torque due to the attachment of the two bollards over 0.06 m to the ropes leads to stretching of the shorter rope and enforces the optical module to never differ more than a few 0.01 m in its positioning [6]. The synthetic ropes are corrosion-free and have minimal creep, and are expected to last 15 years of telescope life. The KM3NeT detector will be deployed at depths greater than 2000 m below the sea surface, which is well below the surface light penetration depth where sharp-toothed predators like sharks that might be a risk to the ropes are not expected to be abundant.

## 3 The ARCA and ORCA detectors of the KM3NeT neutrino telescope

Although similar in construction, two different layouts are designed for the DU-string (Fig. 2) in the two detectors ARCA and ORCA of the KM3NeT telescope. To study astrophysical objects deep in the Universe, Astronomy Research with Cosmics in the Abyss 'ARCA' is planned with 230 DU-strings compacted into two detector building blocks that fill a volume of a cubic kilometre and that will be located at about 3500 m water depth approximately 100 km off Capo Passero, Sicily, Italy. The DU-strings will be 90 m apart horizontally, optical modules separated by 36 m vertically, with the first 70 m above the seafloor not equipped with instrumentation and the total DU-string length about 700 m. A cabled sea-floor network will connect the DU-strings to junction boxes and eventually to a single power and data transmission cable to shore. For this, each DU-string is attached to an anchor which is the interface between the string and the sea-floor network. It carries an electro-optical interlink cable, equipped with wet-mateable connectors and a base container, integrated in the DU-string, that incorporates dedicated optical components and an acoustic receiver used for positioning of the detector elements.

To study neutrino oscillations, optical modules have to be located at shorter distances filling a detector volume of a cubic hectometre. For this, Oscillation Research with Cosmics in the Abyss 'ORCA' is planned with 115 DU-strings in a single building block located at about 2500 m water depth, 40 km south of Toulon, France. The DU-strings will be 20 m apart horizontally, the average vertical distance between optical modules will be 9 m, with the first 30 m above the seafloor not covered by instrumentation, the entire DU-string length totaling about 200 m. As in ARCA, all DU-strings will be interconnected to junction boxes at the seafloor that lead to the power and data transmission cable to shore.

Due to the larger distance between optical modules in ARCA, spacer bars are used between the two ropes that are not used in ORCA. The same deployment method using the LOM is used for both configurations.

## 4 Design of the launching vehicle

### 4.1 General features

Whilst the DU-string's slender design is apt for underwater purposes under current-flow drag, such a mooring cannot be lowered from a ship for deployment without further protective measures. Instead, the entire string is compacted in the 2.18 m diameter aluminium spherical construction frame of the LOM in the laboratory prior to going to sea (Fig. 3), thus allowing for the connection of the VEOC of the DU-string to each of the optical modules under controlled conditions rather than on a moving ship. With the loaded LOM attached to the anchor-frame, the entire package can be lowered to the seafloor using a single heavy ship-winch cable. An acoustic transponder is attached to the acoustic release of the ship-winch cable to locate the anchor-frame to within about 1 m from the designated position. After detachment of the ship's cable, a Remotely Operated underwater Vehicle (ROV) triggers the release of the buoyant LOM from the anchor-weight and thus the unrolling of the DU-string. After completion of the unrolling, the LOM surfaces to be picked up for re-use (Fig. 4).

### 4.2 Launcher vehicle details

The 4.6 kN (2 kN in seawater) weighing LOM is re-usable as it floats to the sea surface after unrolling the DU-string. Inside its seawater-grade aluminium frame, 12 empty 0.44 m diameter glass spheres are mounted to provide a net buoyancy of about 1 kN (Fig. 3). The 1.35 kN top buoy of the DU is housed in the aluminium central shaft of the LOM during deployment. The top buoy is blocked by spring-locks requiring a total force of 0.8 kN for release and two additional spring-loaded locking pins that are released by pulling safety (R-)pins in the end of the Dyneema ropes. The safety pins are pulled by the two ropes after the entire DU-string is unrolled.

The aluminium outer fame of the LOM consists of three circular tracks of two parallel cable trays 0.12 m wide with flanges separated by 120º and 0.05 m high (Fig. 3). The two ropes and the VEOC backbone tube, mechanically attached to one of the ropes, are placed in the cable trays, whereby the tube regularly switches ropes and trays for a balanced string (Fig. 2). Six optical modules are located in cavities between each set of cable trays. Each cavity consists of a carrying frame with guidance flanges and spring-locks (Fig. 5).

An optical module is blocked inside the cavity by two spring-locks holding the bollards and released when the ropes pull with a total force of approximately 0.8 kN. The spring-locks are synchronised by a bracket (Figs 3, 5). The two bollards (one on each side) that connect the collar to the ropes guide each optical module into position via the flanges of a cavity. To prevent the ropes from slacking and potentially hampering the release of an optical module and to correct for rope-length errors, the ropes are tensioned by small springs of about 30 N tension in the flanges (Fig. 5).

The two ropes are connected to a T-bar on the side of the anchor-frame, which weighs approximately 15 kN (Fig. 3). The loaded LOM rests on a circular support structure welded centrally to the anchor-frame. The anchor-frame also serves as a connector between the VEOC backbone of the DU-string and the rest of the cabled seafloor network to the main cable to shore. The anchor-frame and the frame of the LOM are designed to fit in a sea-container for easy transportation.

Alongside the central shaft of the LOM that holds the top buoy, three hollow pipes are mounted from pole to pole (Fig. 3). These pipes are designed to guide cables or slings that attach the loaded LOM to the anchor-frame. Hoisting the complete ensemble of loaded LOM and anchor-weight is thus done by lifting the weight, the cables or slings holding the LOM in its support structure. At the equator of the spherical LOM, three sets of two supports are mounted. Each set of supports can hold the LOM in a rotational spooling traction during the loading of the DU-string in one track of the sphere (Fig. 5).

### 4.3 Custom-made loading tools

The compact loading of a DU-string in the LOM requires a number of custom-made tools. The primary tool is a ‘rotator’ (Fig. 5a), a rotational spooling traction without which loading activities would be rather cumbersome. The rotator holds the LOM in one of three sets of equator-supports. Without a string being loaded, the LOM can be rotated by hand, but the heavier it becomes upon loading, the more the electric motor of the spooling traction is needed for rotation and for holding the LOM at a fixed position for loading-activities like inserting an optical module in its cavity.

A freely rotatable hydraulic ring (partially visible in Fig. 5a under the LOM) is designed such that it is possible to switch between tracks in order to load them all. Assistance is needed of a pallet-carrier for lifting and lowering the ring. This could also be done with a forklift when able to reach fully over the rotator in a 120º relocation, but the positioning of the rotator pins into the equator-supports is more delicate in that case.

When loading, each optical module is put on a manual hydraulic fork-lift with a custom-made ring holding the glass sphere in place (Fig. 5b). The feet of this manual fork-lift are narrow so that the glass-sphere can be driven halfway in the designated cavity of the LOM before being fully pushed inside.

The spring-locks that prevent the optical modules from exiting before being pulled out by the ropes are opened by rotating their synchronising bracket using a custom-made manual tool (Fig. 5c) prior to moving an optical module into its cavity in the LOM. When closed, the spring-locks push the bollards against the end of the flanges thus preventing the optical module to be pushed into the inner shaft of the LOM. A wedge is used to open the additional small springs along the side of the flanges to keep the ropes under slight tension. The top buoy spring-release mechanism is reached from the top of the LOM by approximately 1 m long custom-made hooks (see Section 3.4).

### 4.4 The LOM loading procedure

Prior to coiling it around the LOM, a DU-string is fully integrated with a VEOC, 18 optical modules and a base module. Loading the LOM with the integrated string is done top-down, as the string will be unrolled from the bottom upwards during its launch. The exception is the top buoy, which is loaded last. However, the preparation of its loading is the first step in the loading procedure. To this purpose the ends of the two Dyneema ropes are spliced over equal lengths for connection via two shackles to the buoys' bottom plate later. The two hair-R-pins for unblocking the buoy-release pins are also spliced into the Dyneema ropes and fixed with thin Dyneema line. Thin Dyneema line is also used to attach the prepared rope ends to small holes in the aluminium frame of the LOM using cable-ties, to fix the two rope-ends during the LOM loading rotation. The actual loading is a complex procedure of which details are left out here. Including the mounting on the anchor-weight and all testing, it takes about four days for 2 to 3 trained people to be accomplished. The loading involves continuous back-and-forth reference marks measuring for positioning, making all the precise connections between ropes, VEOC backbone (using several types of clips) and optical modules at the required locations in the LOM.

The two Dyneema ropes are not put on the aluminium frame of the LOM under tension other than hand-tight. The reference marks every 2 m are used to adjust rope length manually, by pulling one of the ropes if needed.

Since the 18 optical modules are divided over three tracks, a track switch takes place every 6 optical modules. During the switch, one rope is given initially 1.5 m more length than the other because of relative positions when taking a turn. The extra length is gained back upon an extra rotation of the LOM. Once also the third track is loaded with optical modules and associated ropes and VEOC backbone, the rope ends are spliced again with equal lengths to within 0.01 m precision to be attached to the anchor-frame.

Finally, the top buoy is mounted into the central shaft of the LOM. The two clamps at the inside of the shaft are opened by pulling them with the two 1-m long hooks for this purpose at the bottom side of the launcher. The buoy is lowered top down into the guidance flanges. The locking pins at its bottom-plate are spring-tensioned for retraction upon deployment with R-pins inserted to avoid this happening immediately (Fig. 6). The long tools are released to close the clamps at the top-side of the buoy. The shackles to which the Dyneema ropes are spliced are fixed to the bottom-plate of the buoy and secured with a cable tie. Now the loaded LOM is ready for mounting on its circular support structure of the anchor-frame.

### 4.5 The anchor-frame release

Whilst previous test ensembles were lowered to the seafloor with a release frame on top of the LOM [7], the self-unrolling is now started by pulling a thin rope once the winch cable is removed. For this release mechanism, the LOM is fixed to a hook on the anchor-frame by a small sling, while two forks at the anchor-frame keep the LOM in the correct orientation. The hook is opened with a small ROV pulling a thin rope (Fig. 7). The release force stays well below 10 N.

## 5. Deployment at sea

The ensemble of the loaded LOM and anchor weighs approximately 26 kN. In seawater the weight is approximately 5.5 kN. The ensemble is lowered with a winch cable to the seafloor from a dedicated ship (Fig. 8). Positioning at the seafloor requires an acoustic positioning system with a transponder to the lowered ensemble. The required accuracy is about ±1 m. Further assistance is needed from an ROV to connect the interlink to the sea-floor network, i.e. connect the VEOC backbone via the base module of the DU-string to the network to shore. After successful connection, of which the verification takes about 30 minutes, the winch cable is released acoustically. This verification entails a full test of the DU-string in furled position: powering on, and testing the communication with each optical module. In principle, the ensemble could be hauled up again to the surface if the DU-string does not pass the verification test. To provide more freedom to the ship's maneuvering and wave motion during the verification, a small buoy is attached a few tens of meters above the LOM to allow for slack in the winch cable without danger of entanglement or collision of the acoustic release with the sphere. Once the release mechanism described in section 3.5 is opened by the ROV pulling the rope near the seafloor the LOM moves upwards from its support structure at the anchor-frame and starts self-unrolling along the two Dyneema ropes.

During the development of the LOM, multiple sea-trials were performed to evaluate and improve the unrolling. The trials were performed with instrumentation for monitoring the translational and rotational speeds of the LOM. A model DU-string, with glass spheres filled with dummy-weights replacing the weight of the optical modules, was recovered each time and repacked in the LOM. As shown in Fig. 9, the unrolling takes 8.7 min for a 680 m string. Initially the speed is about 1.5 m $s^{-1}$ and reduces to about 1.2 m $s^{-1}$ towards the end of the unrolling, slowing down each time an optical module is pulled out and 120 N of buoyancy is lost (Fig. 9b). The slow-down during both track-switches is partially artificial, due to the particular mounting of the accelerometers which provided the data from which the speed is calculated. During the unrolling the LOM moves in a slow precession (Fig. 9c). Due to the mounting of the optical modules and, in ARCA, spacer bars between the ropes, a torque forces the precession of the LOM (and the upright part of the DU-string) back to the desirable upright and parallel position of the two Dyneema ropes. In the trial of Fig. 9c, the back-precession commences after 3.5 min from the start of unrolling. In Fig. 9c also an apparent nutation is observed, especially during the unrolling of the blue track when the LOM is most asymmetrically laden (for track colours see Fig. 3). This nutation has a period of exactly one LOM-turn, and may also be partially due to the way of the accelerometer mounting, which was not perfectly axisymmetric.

The maximum acceleration during pull-out of an optical module is about $2.5g$ [6]. The empty LOM surfaces at a speed of about 1 m $s^{-1}$, and the total unrolling plus surfacing procedure takes about 40 to 50 minutes. The LOM frame, painted for better visibility, is picked up from the surface (Fig. 5). In the trailer of the video 'KM3NeT ORCA line deployment' the deployment of the first ORCA DU-string is shown[1]; two still are in Fig. 8.

## 6. Concluding remarks

After 10 sea-trials, the first complete DU-string deployments were successfully launched with the complete unrolling of an ARCA DU-string up to 750 m long, the undamaged exiting of optical modules and backbone cable and the surfacing of the LOM. At the time of writing, six DU-strings are operating in ORCA and one in ARCA. Loading the LOM and deploying it to the seafloor are procedures that require training of dedicated teams. Taking this into account, the described method may be useful for specific oceanographic mooring deployments, for example the deployment of multiple optical instrumentation like video-monitoring cameras in a string. Such strings may be deployed with or without a backbone.

[1]At https://www.youtube.com/watch?v=dMjN93H7Nvo&list=PLL9OR_-tW5qOtfZigqVpzMmTSwkMjCT1s&index=2&t=0s in the playlist 'Sea Operations' of the KM3NeTneutrino YouTube channel.

**Acknowledgments**

The authors acknowledge the financial support of the funding agencies: Agence Nationale de la Recherche (contract ANR-15-CE31-0020), Centre National de la Recherche Scientifique (CNRS), Commission Européenne (FEDER fund and Marie Curie Program), Institut Universitaire de France (IUF), LabEx UnivEarthS (ANR-10-LABX-0023 and ANR-18-IDEX-0001), Paris Île-de-France Region, France; Shota Rustaveli National Science Foundation of Georgia (SRNSFG, FR-18-1268), Georgia; Deutsche Forschungsgemeinschaft (DFG), Germany; The General Secretariat of Research and Technology (GSRT), Greece; Istituto Nazionale di Fisica Nucleare (INFN), Ministero dell'Università e della Ricerca (MUR), PRIN 2017 program (Grant NAT-NET 2017W4HA7S) Italy; Ministry of Higher Education, Scientific Research and Professional Training, Morocco; Nederlandse organisatie voor Wetenschappelijk Onderzoek (NWO), the Netherlands; The National Science Centre, Poland (2015/18/E/ST2/00758); National Authority for Scientific Research (ANCS), Romania; Ministerio de Ciencia, Innovación, Investigación y Universidades (MCIU): Programa Estatal de Generación de Conocimiento (refs. PGC2018-096663-B-C41, -A-C42, -B-C43, -B-C44) (MCIU/FEDER), Severo Ochoa Centre of Excellence and MultiDark Consolider (MCIU), Junta de Andalucía (ref. SOMM17/6104/UGR), Generalitat Valenciana: Grisolía (ref. GRISOLIA/2018/119) and GenT (ref. CIDEGENT/2018/034) programs, La Caixa Foundation (ref. LCF/BQ/IN17/11620019), EU: MSC program (ref. 713673), Spain.

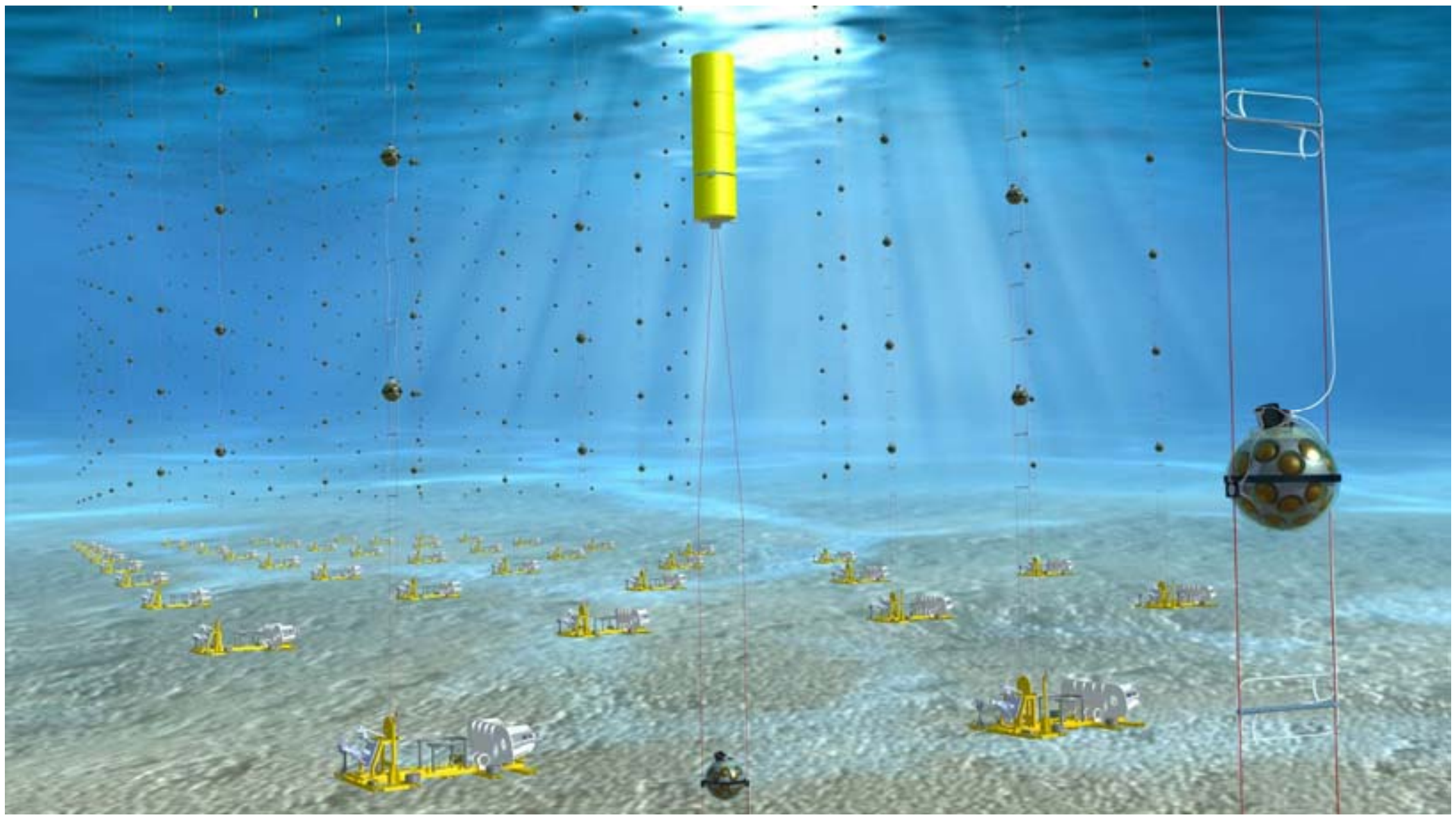

**Figure 1**. Artist impression of the finalised KM3NeT underwater neutrino telescope, with a single optical module held in a (double-rope) string in the right-hand-side foreground. The white tube along the ropes is the oil-filled backbone with copper wire and optical fibres inside. The yellow cylinder in the centre of the image is the top buoy, which is above each string.

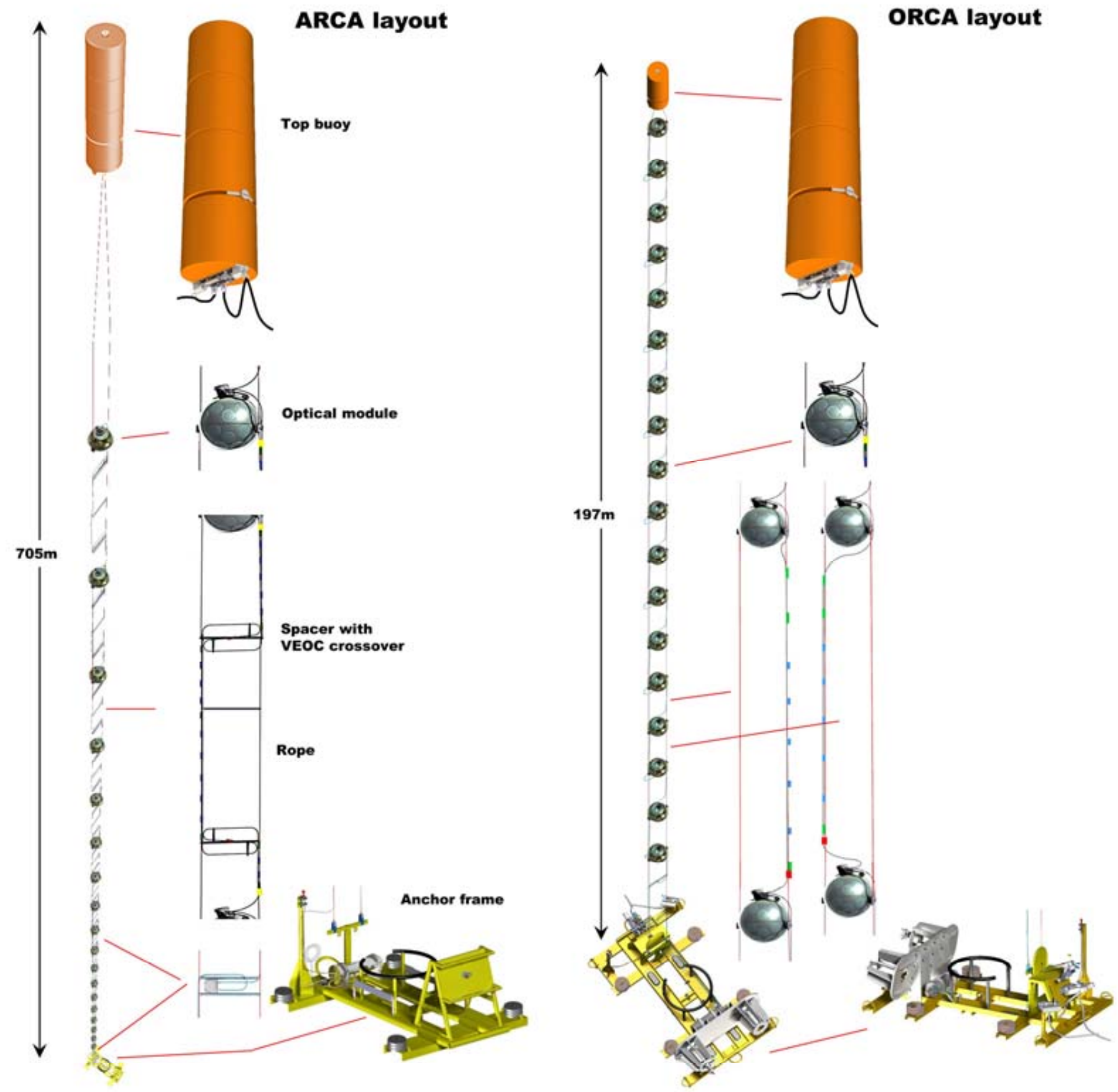

**Figure 2**. ARCA (left) and ORCA (right) DU-string layouts showing the 18 optical modules on one DU-string, the VEOC guided along the DU-string, as well as the top buoys and anchor frames.

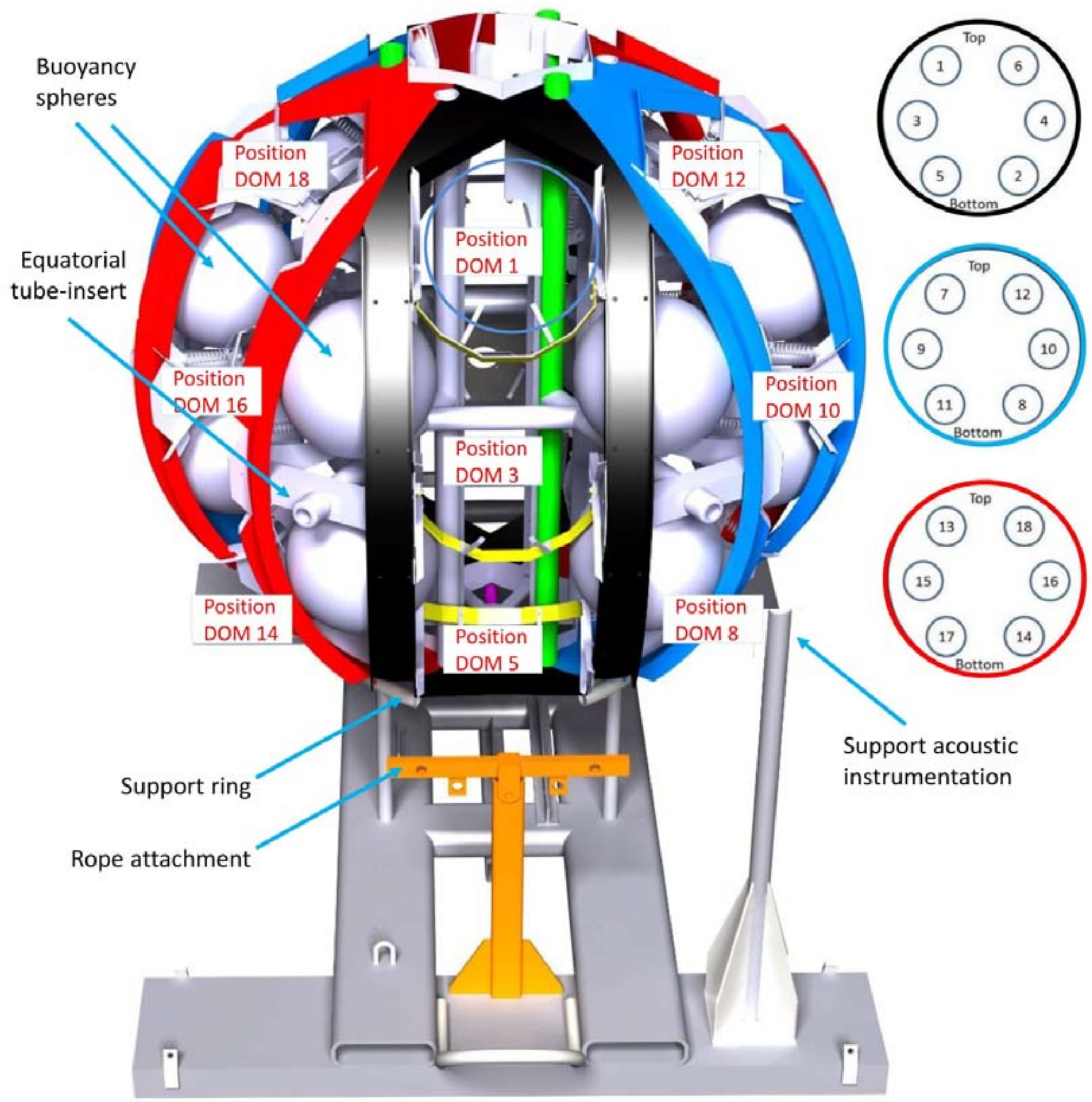

**Figure 3**. Artist impression of the LOM on its circular support structure of the anchor-frame. In black, red and blue the three LOM tracks with open slots for optical modules (labelled DOM 1,…,18). To the right, the order of packing optical modules is shown per circular track (ORCA version). In green, the three pole-to-pole shafts for the slings to hoist the ensemble of LOM and anchor-frame are shown. In the foreground, the orange T-bar for attaching the two Dyneema ropes to the anchor is visible. The shaft to the right-foreground is for mounting the hydrophone for acoustic positioning purposes.

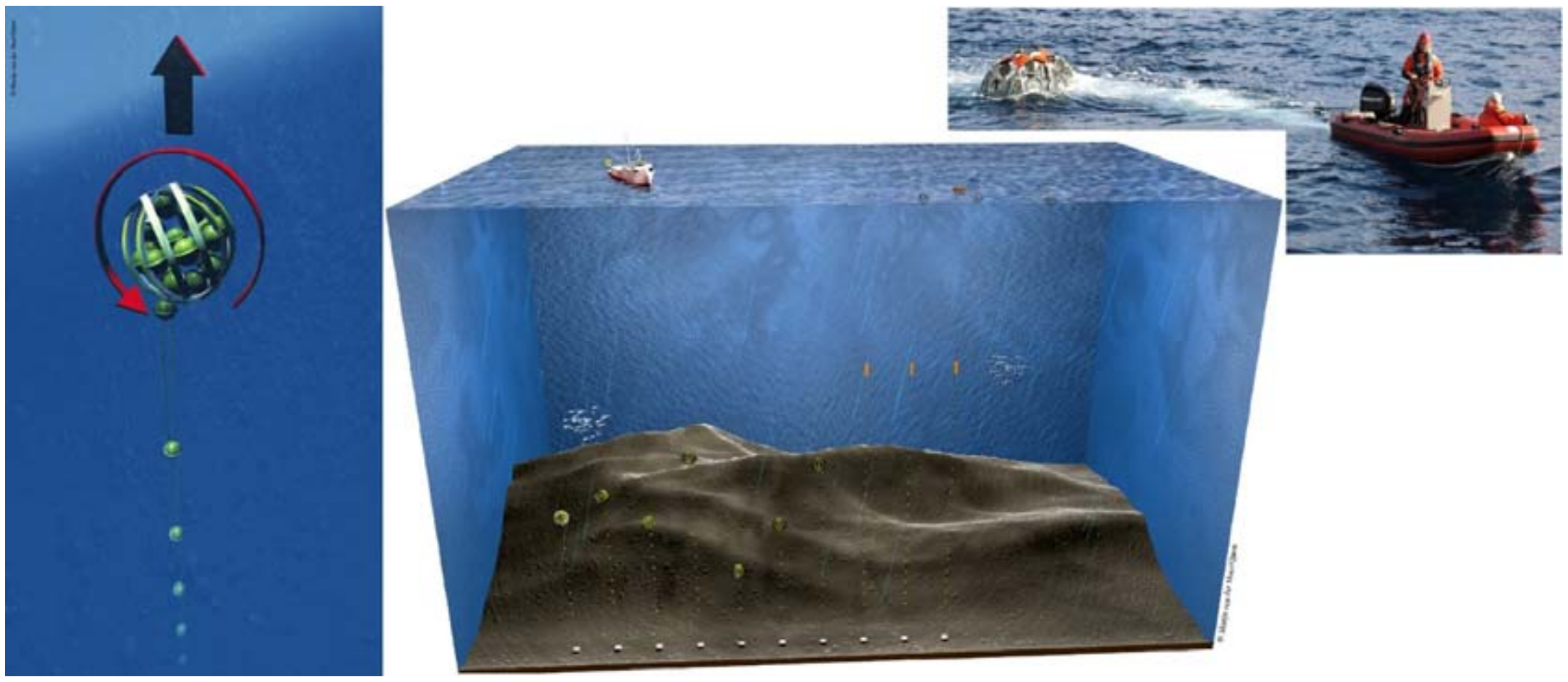

**Figure 4**. Artist impression of the unrolling of a detection unit made by Marijn van der Meer/Quest. Under Creative Common: CC BY-NC 4.0, https://creativecommons.org/licences/by-nc/4.0. The LOM is recovered from the surface after successful unrolling. In reality deployment of DU-strings is done one after another, not quasi-simultaneously.

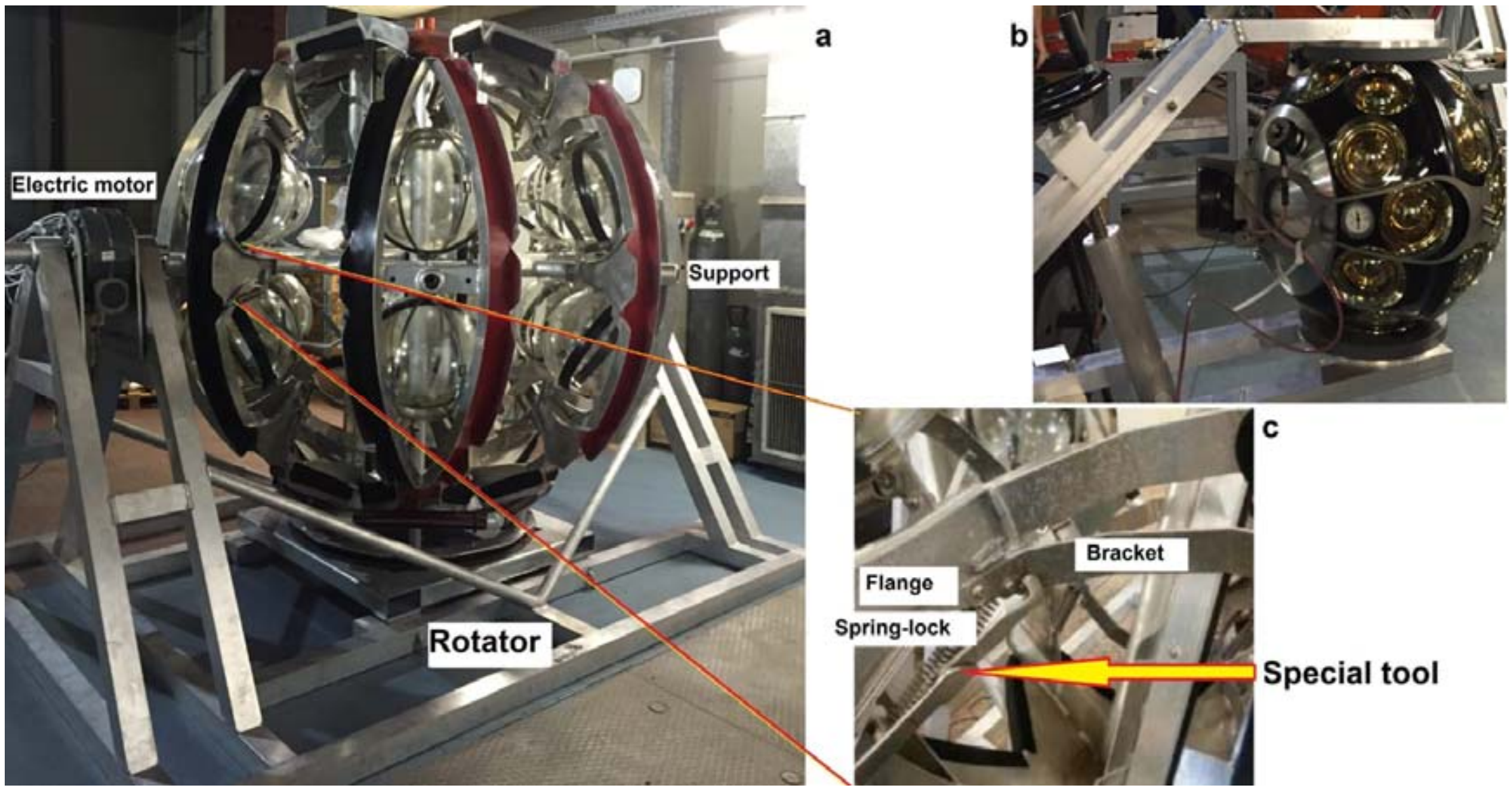

**Figure 5**. Loading of the LOM with a DU-string. (a) LOM mounted on the rotator frame. Below the LOM, the ring for changing tracks is partially visible. (b) Detail of custom-made forklift tool holding an optical module for placing it in an open cavity of the LOM. (c) Spring-lock for positioning and holding the optical module in its cavity, together with its special tool for opening the lock in the two flanges by rotating their synchronising bracket. In (c) the lock is opened, in (a) all locks are closed.

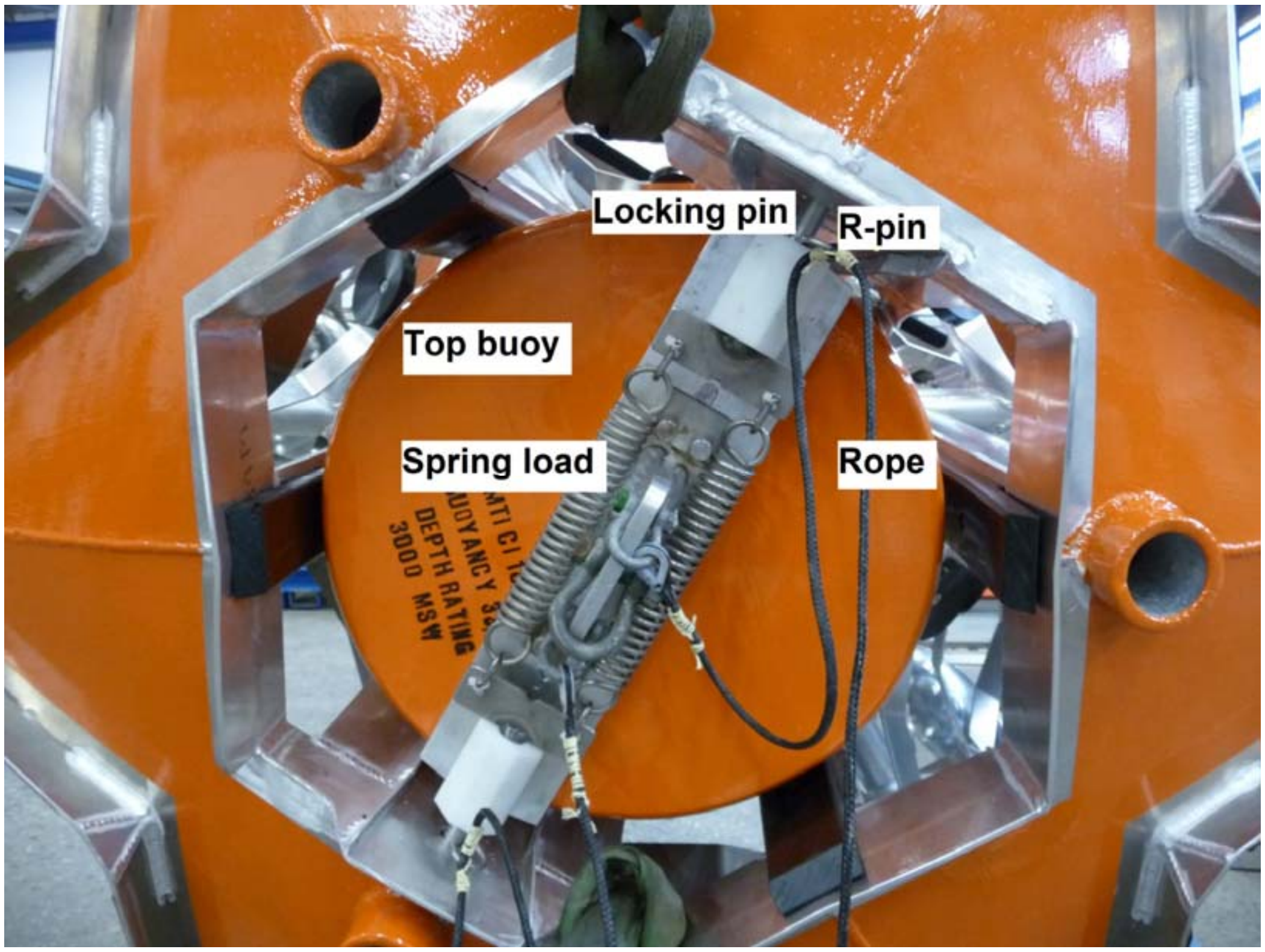

**Figure 6**. LOM viewed from above, facing the bottom plate of the top buoy and its release mechanism.

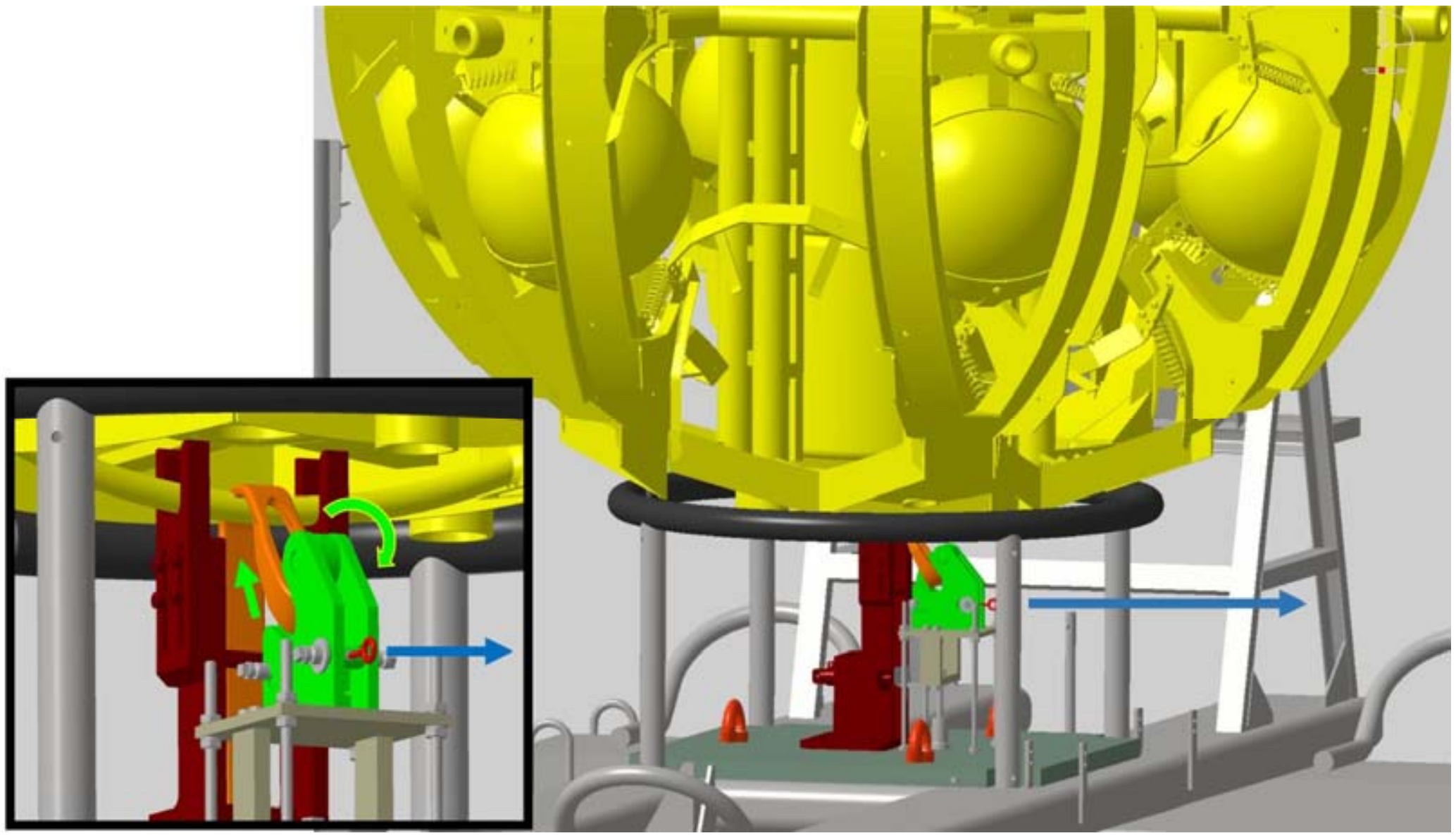

**Figure 7**. Anchor-frame release system. The long arrow pointing to the right shows the direction of pull by the ROV. After ROV-pull, the sling is released on one side and the sphere is free to move up and start the unrolling.

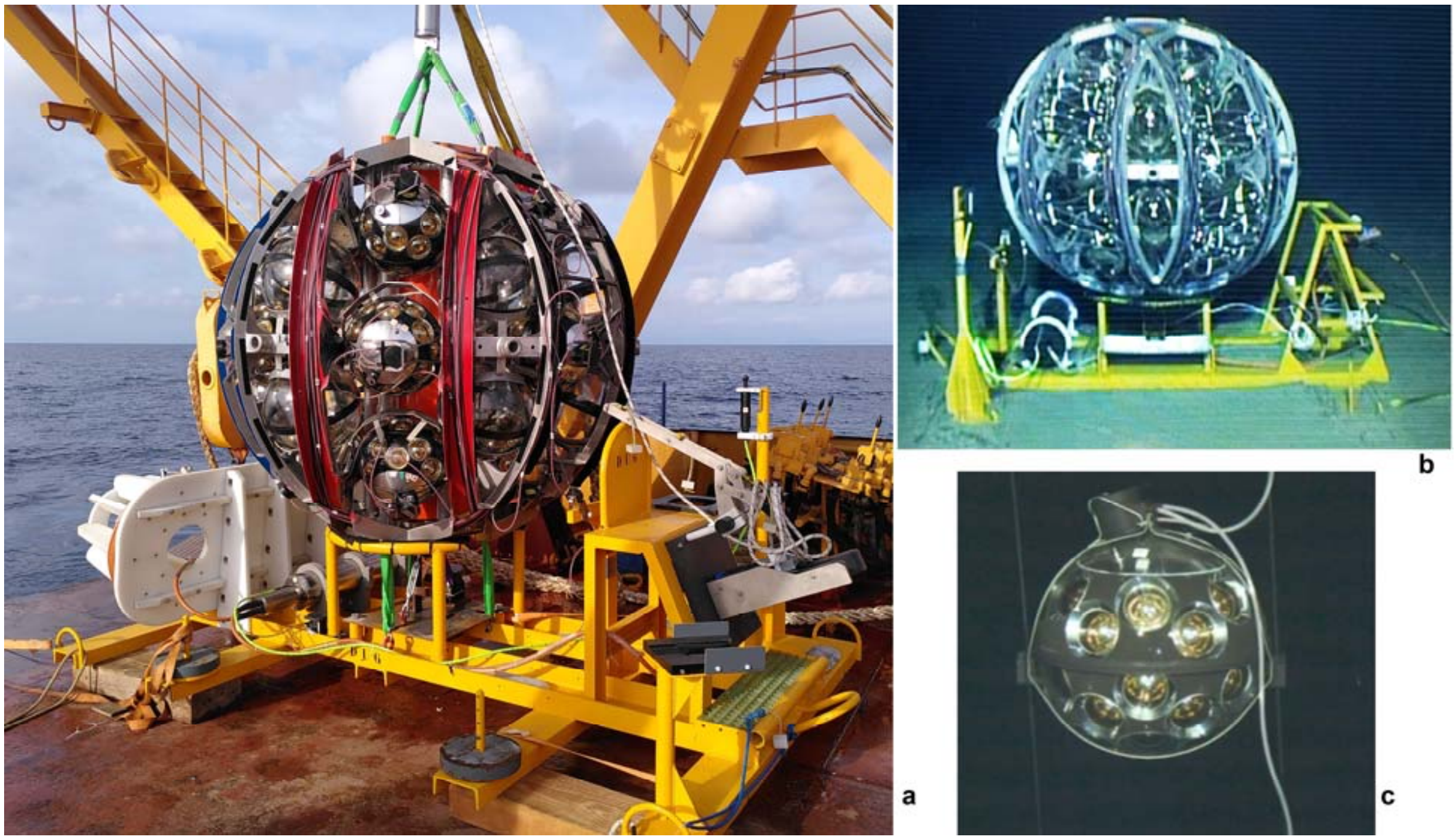

**Figure 8**. (a) Fully loaded LOM on its anchor-frame ready for deployment. (b) ROV image of successful deployment of first DU-string, with the loaded LOM at the seafloor just before unrolling. (c) ROV image of a single optical module after unrolling.

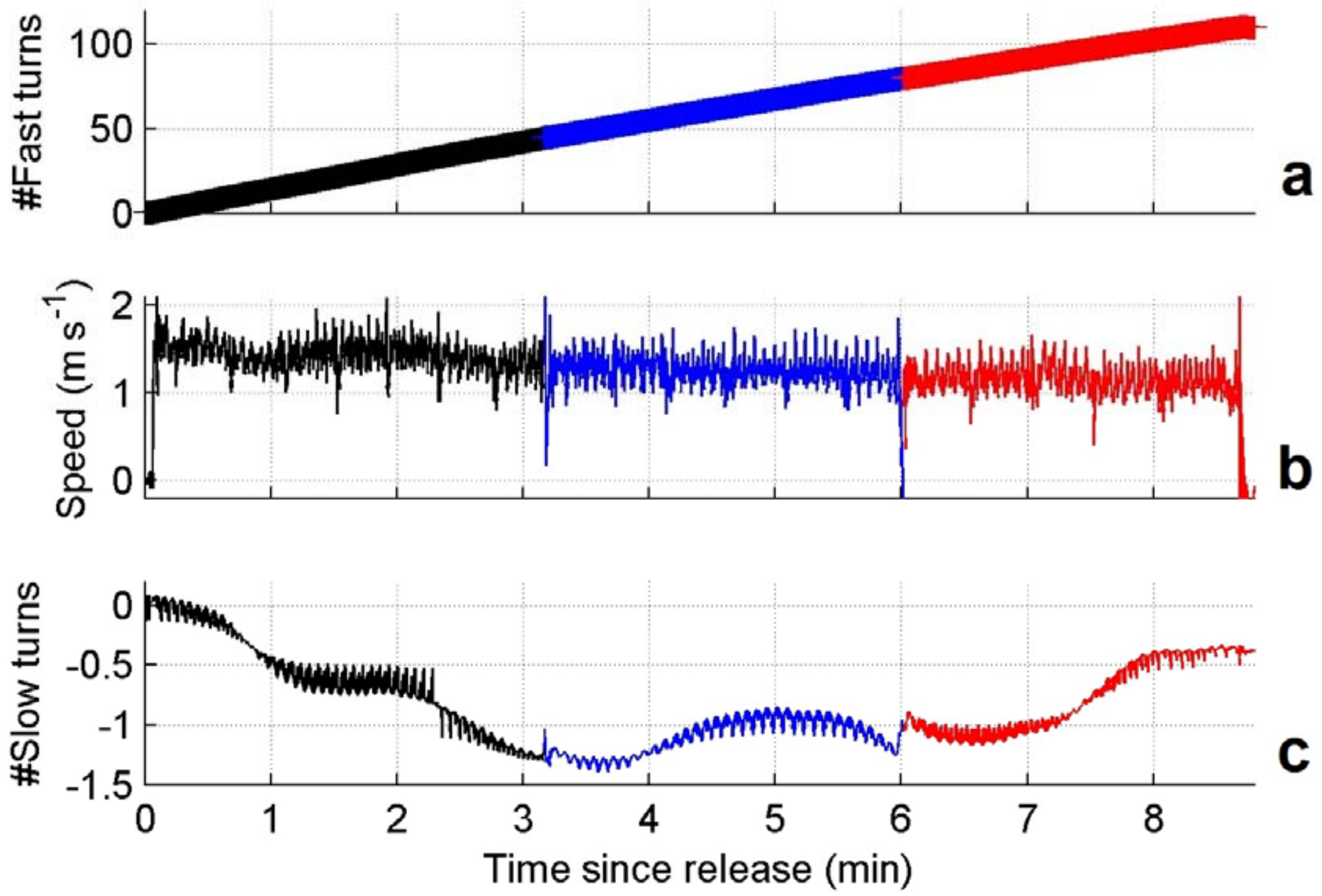

**Figure 9.** Time series of self-unrolling of the DU-string from the LOM frame during a sea-trial. The three colours associate with the three tracks on the LOM (cf. Fig. 3). (a) Number of 6.3 m (LOM circumference) turns. (b) Approximate vertical translation speed, computed from the data in (a). (c) Number of precession turns.